\documentclass[10pt,a4paper,english,twocolumn, conference]{IEEEtran}
\usepackage[T1]{fontenc}
\usepackage{color}
\usepackage{verbatim}
\usepackage{float}
\usepackage{textcomp}
\usepackage{mathrsfs}
\usepackage{amsmath}
\usepackage{amsthm}
\usepackage{graphicx}
\usepackage{setspace}

\makeatletter

\pdfpageheight\paperheight
\pdfpagewidth\paperwidth

\floatstyle{ruled}
\newfloat{algorithm}{tbp}{loa}
\providecommand{\algorithmname}{Algorithm}
\floatname{algorithm}{\protect\algorithmname}

\theoremstyle{plain}
\newtheorem{thm}{\protect\theoremname}

\usepackage{subfigure}
\usepackage{epstopdf}
\usepackage{cite}
\usepackage{citesort}
\usepackage{balance}

\makeatother

\usepackage{babel}
\providecommand{\theoremname}{Theorem}

\begin{document}

\title{Please Lower Small Cell Antenna Heights in 5G
}

\author{\noindent {\normalsize{}Ming Ding, }\emph{\normalsize{}Data61, Australia}{\normalsize{}
\{Ming.Ding@data61.csiro.au\}}\\
{\normalsize{}David L$\acute{\textrm{o}}$pez P$\acute{\textrm{e}}$rez,
}\emph{\normalsize{}Nokia Bell Labs, Ireland}{\normalsize{} \{david.lopez-perez@nokia.com\}}\textit{\footnotesize{}}\\
}
\maketitle
\begin{abstract}
In this paper, we present a new and significant theoretical discovery.
If the absolute height difference between base station (BS) antenna
and user equipment (UE) antenna is larger than zero, then the network
capacity performance in terms of the area spectral efficiency (ASE)
will \emph{continuously decrease} as the BS density increases for
ultra-dense (UD) small cell networks (SCNs). This performance behavior
has a tremendous impact on the deployment of UD SCNs in the 5th-generation
(5G) era. Network operators may invest large amounts of money in deploying
more network infrastructure to only obtain an even worse network performance.
Our study results reveal that it is a must to lower the SCN BS antenna
height to the UE antenna height to fully achieve the capacity gains
of UD SCNs in 5G. However, this requires a revolutionized approach
of BS architecture and deployment, which is explored in this paper
too.

\footnote{1536-1276 © 2015 IEEE. Personal use is permitted, but republication/redistribution requires IEEE permission. Please find the final version in IEEE from the link: http://ieeexplore.ieee.org/document/7842150/. Digital Object Identifier: 10.1109/GLOCOM.2016.7842150}

\end{abstract}

\section{Introduction\label{sec:Introduction}}

From 1950 to 2000, the wireless network capacity has increased around
1 million fold, in which an astounding 2700\texttimes{} gain was achieved
through network densification using smaller cells~\cite{Webb_survey}.
After 2008, network densification continues to fuel the 3rd Generation
Partnership Project (3GPP) 4th-generation (4G) Long Term Evolution
(LTE) networks, and is expected to remain as one of the main forces
to drive the 5th-generation (5G) networks onward~\cite{Tutor_smallcell}.
Indeed, the orthogonal deployment of ultra-dense (UD) small cell networks
(SCNs) within the existing macrocell network, i.e., small cells and
macrocells operating on different frequency spectrum (3GPP Small Cell
Scenario \#2a~\cite{TR36.872}), is envisaged as the workhorse for
capacity enhancement in 5G due to its large spectrum reuse and its
easy management; the latter one arising from its low interaction with
the macrocell tier, e.g., no inter-tier interference~\cite{Tutor_smallcell}.
In this paper, the focus is on the analysis of these UD SCNs with
an orthogonal deployment with the macrocells.

Before 2015, the common understanding on SCNs was that the density
of base stations (BSs) would not affect the per-BS coverage probability
performance in interference-limited fully-loaded wireless networks,
and thus the area spectral efficiency (ASE) performance in $\textrm{bps/Hz/km}^{2}$
would scale linearly with network densification~\cite{Jeff2011}.
The implication of such conclusion is huge: %
\emph{The BS density does NOT matter}, since the increase in the interference
power caused by a denser network would be exactly compensated by the
increase in the signal power due to the reduced distance between transmitters
and receivers. Fig.~\ref{fig:comp_ASE_2Gto5G} shows the theoretical
ASE performance predicted in~\cite{Jeff2011}.%
{} However, it is important to note that this conclusion was obtained
with considerable simplifications on the propagation environment,
which should be placed under scrutiny when evaluating dense and UD
SCNs, since they are fundamentally different from sparse ones in various
aspects~\cite{Tutor_smallcell}.

\noindent \begin{center}
\begin{figure}
\noindent \begin{centering}
\includegraphics[width=8.2cm]{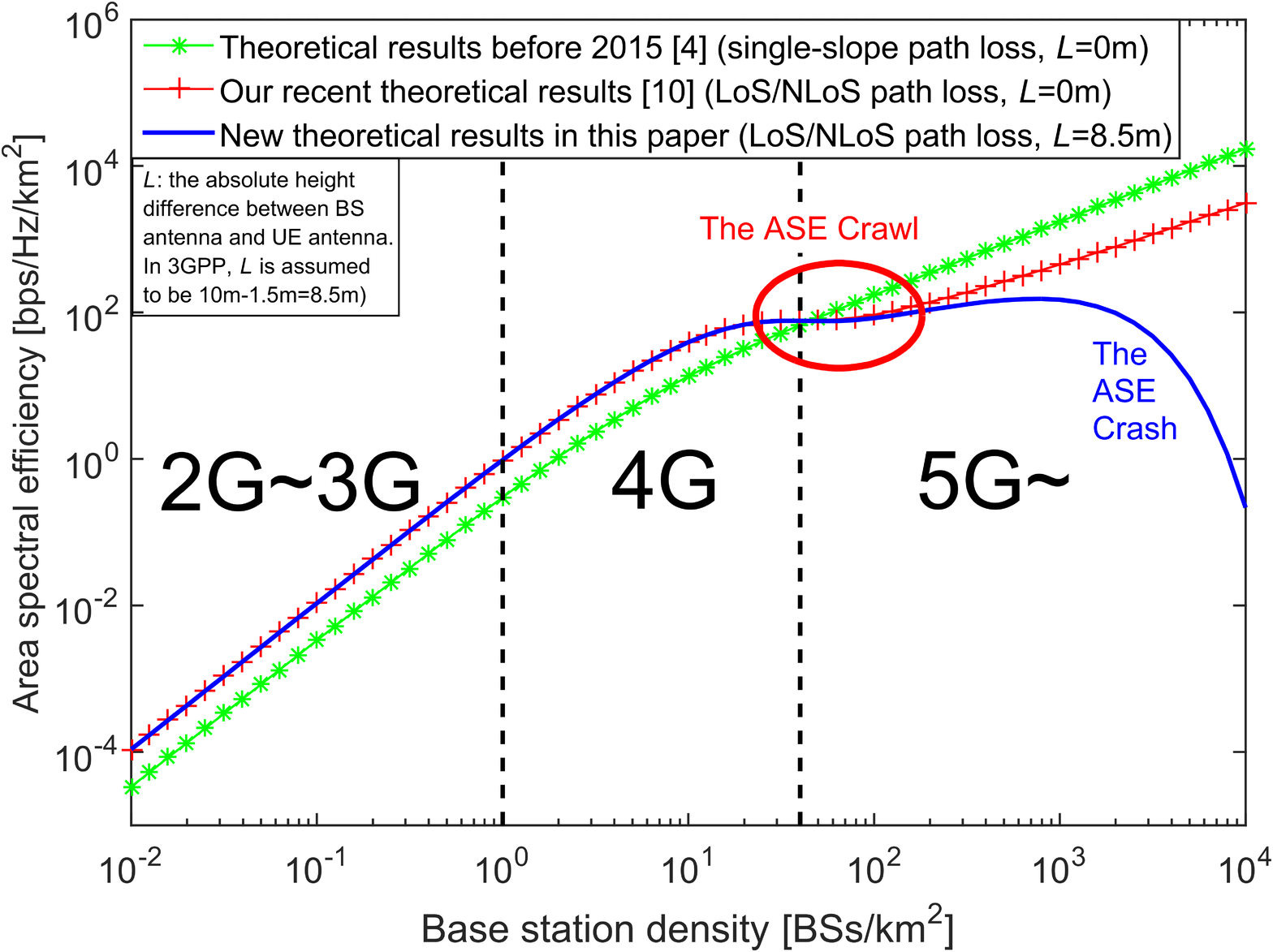}\renewcommand{\figurename}{Fig.}\caption{\label{fig:comp_ASE_2Gto5G}Theoretical comparison of the ASE performance
in $\textrm{bps/Hz/km}^{2}$. Note that all the results are obtained
using practical 3GPP channel models~\cite{TR36.828,SCM_pathloss_model},
which will be explained in details later. Due to the practicality
of the used channel models, the results shown here accurately characterize
realistic telecommunication systems both qualitatively and quantitatively.
For example, considering a typical bandwidth of 10\,MHz\textasciitilde{}100\,MHz
for the state-of-the-art LTE network, the achievable area throughput
is in the order of several $\textrm{Gbps/km}^{2}$, because the ASE
for 4G is shown to be around 100\,$\textrm{bps/Hz/km}^{2}$. }
\par\end{centering}
\vspace{-0.5cm}
\end{figure}
\par\end{center}

\vspace{-0.8cm}

In the last year, a few noteworthy studies have been carried out%
{} to revisit the network performance analysis for UD SCNs under %
more practical propagation assumptions. In~\cite{related_work_Jeff},
the authors considered a multi-slope piece-wise path loss function,
while in~\cite{Related_work_Health}, the authors investigated line-of-sight
(LoS) and non-line-of-sight (NLoS) transmission as a probabilistic
event for a millimeter wave communication scenario. The most important
finding in these two works was that the per-BS coverage probability
performance starts to decrease when the BS density is sufficiently
large. Fortunately, such decrease of coverage probability did not
change the monotonic increase of the ASE as the BS density increases.

In our very recent work~\cite{our_GC_paper_2015_HPPP,our_work_TWC2016},
we took a step further and generalized the works in~\cite{related_work_Jeff}
and~\cite{Related_work_Health} by considering both piece-wise path
loss functions and probabilistic NLoS and LoS transmissions. Our new
finding was not only quantitatively but also qualitatively different
from previous results in~\cite{Jeff2011,related_work_Jeff,Related_work_Health}:
The ASE will suffer from a slow growth or even a small\emph{ decrease}
on the journey from 4G to 5G when the BS density is larger than a
threshold. Fig.~\ref{fig:comp_ASE_2Gto5G} shows these new theoretical
results on the ASE performance, where such threshold is around 20\,$\textrm{BSs/km}^{2}$
and the slow/negative ASE growth is highlighted by a circled area.
This circled area is referred to as \emph{the ASE Crawl} hereafter.
The intuition of \emph{the ASE Crawl} is that the interference power
increases faster than the signal power due to the transition of a
large number of interference paths from NLoS to LoS with the network
densification. The implication is profound:%
\emph{ The BS density DOES matter,} since it affects the signal to
interference relationship. Thus, operators should be careful when
deploying dense SCNs in order to avoid investing huge amounts of money
and end up obtaining an even worse network performance due to \emph{the
ASE Crawl}. %
Fortunately, our results in~\cite{our_GC_paper_2015_HPPP,our_work_TWC2016}
also pointed out that the ASE will again grow almost linearly as the
network further evolves to an UD one%
, i.e., $>10^{3}\thinspace\textrm{BSs/km}^{2}$ in Fig.~\ref{fig:comp_ASE_2Gto5G}.
According to our results and considering a 300\,MHz bandwidth, if
the BS density can go as high as $10^{4}\thinspace\textrm{BSs/km}^{2}$,
the problem of \emph{the ASE Crawl} %
caused by the NLoS to LoS transition can be overcome, and an area
throughput of $10^{3}\thinspace\textrm{Gbps/km}^{2}$ can be achieved,
thus opening up an efficient way forward to 5G.

Unfortunately, the NLoS to LoS transition is not the only obstacle
to efficient UD SCNs in 5G, and there are more challenges to overcome
to get there. In this paper, we present for the first time the serious
problem posed by the absolute antenna height difference between SCN
base stations (BSs) and user equipments (UEs), and evaluate its impact
on UD SCNs by means of a three-dimensional (3D) stochastic geometry
analysis (SGA). We made a new and significant theoretical discovery:
\textbf{If the absolute antenna height difference between BSs and
UEs, denoted by $L$, is larger than zero, then the ASE performance
will }\textbf{\emph{continuously decrease}}\textbf{ as the SCN goes
ultra-dense.} Fig.~\ref{fig:comp_ASE_2Gto5G} illustrates the significance
of such theoretical finding with $L=8.5$\,m~\cite{TR36.814}: After
\emph{the ASE Crawl}, the ASE performance only increases marginally
(\textasciitilde{}1.4x) from 109.1\,$\textrm{bps/Hz/km}^{2}$ to
149.6\,$\textrm{bps/Hz/km}^{2}$ as the BS density goes from $200\thinspace\textrm{BSs/km}^{2}$
to $10^{3}\thinspace\textrm{BSs/km}^{2}$%
, which is then followed by a continuous and quick fall starting from
around $10^{3}\thinspace\textrm{BSs/km}^{2}$. The implication of
this result is even more profound than that of \emph{the ASE Crawl},
since following a traditional deployment with UD SCN BSs deployed
at lamp posts or similar heights may dramatically reduce the network
performance in 5G%
. Such decline of ASE in UD SCNs will be referred to as \emph{the
ASE Crash} hereafter, and its fundamental reasons will be explained
in details later in this paper.

In order to address this serious problem of \emph{the ASE Crash},
we further propose to change the traditional BS deployment, and lower
the 5G UD SCN BS antenna height %
to the UE antenna height, so that the ASE behavior of UD SCNs can
roll back to our previous results in~\cite{our_work_TWC2016}, thus
avoiding \emph{the ASE Crash. }%
This %
requires a revolutionized BS deployment approach, which will also
be explored in this paper.

The rest of this paper is structured as follows. Section~\ref{sec:System-Model}
describes the system model for the 3D SGA. Section~\ref{sec:General-Results}
presents our theoretical results on the coverage probability and the
ASE performance%
, while the numerical results are discussed in Section~\ref{sec:Simulation-and-Discussion},
with remarks shedding new light on the revolutionized BS deployment
with UE-height antennas. Finally, the conclusions are drawn in Section~\ref{sec:Conclusion}.

\section{System Model\label{sec:System-Model}}

We consider a downlink (DL) cellular network with BSs deployed on
a plane according to a homogeneous Poisson point process (HPPP) $\Phi$
of intensity $\lambda$ $\textrm{BSs/km}^{2}$%
. UEs are Poisson distributed in the considered network with an intensity
of $\rho$ $\textrm{UEs/km}^{2}$. Note that $\rho$ is assumed to
be sufficiently larger than $\lambda$ so that each BS has at least
one associated UE in its coverage~\cite{related_work_Jeff,Related_work_Health,our_GC_paper_2015_HPPP,our_work_TWC2016}.
The two-dimensional (2D) distance between a BS and an a UE is denoted
by $r$%
. Moreover, the absolute antenna height difference between a BS and
a UE is %
denoted by $L$%
. Hence, the 3D distance between a BS and a UE%
{} can be expressed as

\begin{singlespace}
\noindent
\begin{equation}
w=\sqrt{r^{2}+L^{2}}.\label{eq:actual_dis_BS2UE}
\end{equation}

\end{singlespace}

\vspace{-0.1cm}
Following~\cite{our_GC_paper_2015_HPPP,our_work_TWC2016}, we adopt
a very general and practical path loss model, in which the path loss
$\zeta\left(w\right)$ associated with distance $w$ is segmented
into $N$ pieces written as%

\begin{singlespace}
\noindent
\begin{equation}
\zeta\left(w\right)=\begin{cases}
\zeta_{1}\left(w\right), & \textrm{when }0\leq w\leq d_{1}\\
\zeta_{2}\left(w\right), & \textrm{when }d_{1}<w\leq d_{2}\\
\vdots & \vdots\\
\zeta_{N}\left(w\right), & \textrm{when }w>d_{N-1}
\end{cases},\label{eq:prop_PL_model}
\end{equation}

\end{singlespace}

\vspace{-0.1cm}

\noindent where each piece $\zeta_{n}\left(w\right),n\in\left\{ 1,2,\ldots,N\right\} $
is modeled as

\noindent
\begin{equation}
\zeta_{n}\left(w\right)\hspace{-0.1cm}=\hspace{-0.1cm}\begin{cases}
\hspace{-0.2cm}\begin{array}{l}
\zeta_{n}^{\textrm{L}}\left(w\right)=A_{n}^{{\rm {L}}}w^{-\alpha_{n}^{{\rm {L}}}},\\
\zeta_{n}^{\textrm{NL}}\left(w\right)=A_{n}^{{\rm {NL}}}w^{-\alpha_{n}^{{\rm {NL}}}},
\end{array} & \hspace{-0.2cm}\hspace{-0.3cm}\begin{array}{l}
\textrm{LoS:}~\textrm{Pr}_{n}^{\textrm{L}}\left(w\right)\\
\textrm{NLoS:}~1-\textrm{Pr}_{n}^{\textrm{L}}\left(w\right)
\end{array}\hspace{-0.1cm},\end{cases}\label{eq:PL_BS2UE}
\end{equation}

\noindent where $\zeta_{n}^{\textrm{L}}\left(w\right)$ and $\zeta_{n}^{\textrm{NL}}\left(w\right),n\in\left\{ 1,2,\ldots,N\right\} $
are the $n$-th piece path loss functions for the LoS transmission
and the NLoS transmission, respectively, $A_{n}^{{\rm {L}}}$ and
$A_{n}^{{\rm {NL}}}$ are the path losses at a reference distance
$w=1$ for the LoS and the NLoS cases, respectively, and $\alpha_{n}^{{\rm {L}}}$
and $\alpha_{n}^{{\rm {NL}}}$ are the path loss exponents for the
LoS and the NLoS cases, respectively. In practice, $A_{n}^{{\rm {L}}}$,
$A_{n}^{{\rm {NL}}}$, $\alpha_{n}^{{\rm {L}}}$ and $\alpha_{n}^{{\rm {NL}}}$
are constants obtainable from field tests~\cite{TR36.828,SCM_pathloss_model}.%
{} Moreover, $\textrm{Pr}_{n}^{\textrm{L}}\left(w\right)$ is the $n$-th
piece LoS probability function that a transmitter and a receiver separated
by a distance $w$ has a LoS path, which is assumed to be a monotonically
decreasing function with regard to $w$.

For convenience, $\left\{ \zeta_{n}^{\textrm{L}}\left(w\right)\right\} $
and $\left\{ \zeta_{n}^{\textrm{NL}}\left(w\right)\right\} $ are
further stacked into piece-wise functions written as

\begin{singlespace}
\noindent
\begin{equation}
\zeta^{Path}\left(w\right)=\begin{cases}
\zeta_{1}^{Path}\left(w\right), & \textrm{when }0\leq w\leq d_{1}\\
\zeta_{2}^{Path}\left(w\right),\hspace{-0.3cm} & \textrm{when }d_{1}<w\leq d_{2}\\
\vdots & \vdots\\
\zeta_{N}^{Path}\left(w\right), & \textrm{when }w>d_{N-1}
\end{cases},\label{eq:general_PL_func}
\end{equation}

\end{singlespace}

\noindent where the string variable $Path$ takes the value of ``L''
and ``NL'' for the LoS and the NLoS cases, respectively.

Besides, $\left\{ \textrm{Pr}_{n}^{\textrm{L}}\left(w\right)\right\} $
is stacked into a piece-wise function as

\begin{singlespace}
\noindent
\begin{equation}
\textrm{Pr}^{\textrm{L}}\left(w\right)=\begin{cases}
\textrm{Pr}_{1}^{\textrm{L}}\left(w\right), & \textrm{when }0\leq w\leq d_{1}\\
\textrm{Pr}_{2}^{\textrm{L}}\left(w\right),\hspace{-0.3cm} & \textrm{when }d_{1}<w\leq d_{2}\\
\vdots & \vdots\\
\textrm{Pr}_{N}^{\textrm{L}}\left(w\right), & \textrm{when }w>d_{N-1}
\end{cases}.\label{eq:general_LoS_Pr}
\end{equation}

\end{singlespace}

In this paper, we assume a practical user association strategy (UAS),
in which each UE should be associated with the BS providing the smallest
path loss (i.e., with the largest $\zeta\left(w\right)$)~\cite{Related_work_Health,our_work_TWC2016}.
In addition, we assume that each BS/UE is equipped with an isotropic
antenna, and that the multi-path fading between a BS and a UE is modeled
as independently identical distributed (i.i.d.) Rayleigh fading~\cite{related_work_Jeff,Related_work_Health,our_GC_paper_2015_HPPP,our_work_TWC2016}.
Note that a more practical Rician fading will also be considered in
the simulation section to show its impact on our conclusions.

\section{Main Results\label{sec:General-Results}}

Using a 3D SGA based on the HPPP theory, we study the performance
of the SCN by considering the performance of a typical UE located
at the origin $o$.

We first investigate the coverage probability that %
this UE's signal-to-interference-plus-noise ratio (SINR) is above
a per-designated threshold $\gamma$:

\begin{singlespace}
\noindent
\begin{equation}
p^{\textrm{cov}}\left(\lambda,\gamma\right)=\textrm{Pr}\left[\mathrm{SINR}>\gamma\right],\label{eq:Coverage_Prob_def}
\end{equation}

\end{singlespace}

\noindent where the SINR is calculated as

\vspace{0.2cm}

\noindent
\begin{equation}
\mathrm{SINR}=\frac{P\zeta\left(w\right)h}{I_{\textrm{agg}}+N_{0}},\label{eq:SINR}
\end{equation}

\noindent where $h$ is the channel gain and is modeled as an exponential
random variable (RV) with the mean of one due to%
{} Rayleigh fading%
, $P$ and $N_{0}$ are the transmission power of each BS and the
additive white Gaussian noise (AWGN) power at each UE, respectively,
and $I_{\textrm{agg}}$ is the cumulative interference given by

\noindent
\begin{equation}
I_{\textrm{agg}}=\sum_{i:\,b_{i}\in\Phi\setminus b_{o}}P\beta_{i}g_{i},\label{eq:cumulative_interference}
\end{equation}

\noindent where $b_{o}$ is the BS serving the typical UE located
at distance $w$ from the typical UE, and $b_{i}$, $\beta_{i}$ and
$g_{i}$ are the $i$-th interfering BS, the path loss associated
with $b_{i}$ and the multi-path fading channel gain associated with
$b_{i}$, respectively.

Based on the path loss model in (\ref{eq:prop_PL_model}) with 3D
distances and the considered UAS, we present our main result on $p^{\textrm{cov}}\left(\lambda,\gamma\right)$
in Theorem~\ref{thm:p_cov_UAS1} shown on the top of the next page.

\noindent
\begin{algorithm*}[!tp]
\begin{thm}
{\small{}\label{thm:p_cov_UAS1}Considering the path loss model in
(\ref{eq:prop_PL_model}) and }the presented UAS{\small{}, the probability
of coverage $p^{{\rm {cov}}}\left(\lambda,\gamma\right)$ can be derived
as}{\small \par}

\noindent {\small{}
\begin{equation}
p^{{\rm {cov}}}\left(\lambda,\gamma\right)=\sum_{n=1}^{N}\left(T_{n}^{{\rm {L}}}+T_{n}^{{\rm {NL}}}\right),\label{eq:Theorem_1_p_cov}
\end{equation}
}{\small \par}

\noindent \vspace{-0.1cm}

\noindent {\small{}where $T_{n}^{{\rm {L}}}=\int_{\sqrt{d_{n-1}^{2}-L^{2}}}^{\sqrt{d_{n}^{2}-L^{2}}}{\rm {Pr}}\left[\frac{P\zeta_{n}^{{\rm {L}}}\left(\sqrt{r^{2}+L^{2}}\right)h}{I_{{\rm {agg}}}+N_{0}}>\gamma\right]f_{R,n}^{{\rm {L}}}\left(r\right)dr$,
$T_{n}^{{\rm {NL}}}=\int_{\sqrt{d_{n-1}^{2}-L^{2}}}^{\sqrt{d_{n}^{2}-L^{2}}}{\rm {Pr}}\left[\frac{P\zeta_{n}^{{\rm {NL}}}\left(\sqrt{r^{2}+L^{2}}\right)h}{I_{{\rm {agg}}}+N_{0}}>\gamma\right]f_{R,n}^{{\rm {NL}}}\left(r\right)dr$,
and $d_{0}$ and $d_{N}$ are defined as $L$ and $+\infty$, respectively.
Moreover, $f_{R,n}^{{\rm {L}}}\left(r\right)$ and $f_{R,n}^{{\rm {NL}}}\left(r\right)$
$\left(\sqrt{d_{n-1}^{2}-L^{2}}<r\leq\sqrt{d_{n}^{2}-L^{2}}\right)$,
are represented by}{\small \par}

\noindent {\small{}
\begin{equation}
f_{R,n}^{{\rm {L}}}\left(r\right)=\exp\left(\hspace{-0.1cm}-\hspace{-0.1cm}\int_{0}^{r_{1}}\left(1-{\rm {Pr}}^{{\rm {L}}}\left(\sqrt{u^{2}+L^{2}}\right)\right)2\pi u\lambda du\right)\exp\left(\hspace{-0.1cm}-\hspace{-0.1cm}\int_{0}^{r}{\rm {Pr}}^{{\rm {L}}}\left(\sqrt{u^{2}+L^{2}}\right)2\pi u\lambda du\right){\rm {Pr}}_{n}^{{\rm {L}}}\left(\sqrt{r^{2}+L^{2}}\right)2\pi r\lambda,\hspace{-0.1cm}\hspace{-0.1cm}\hspace{-0.1cm}\hspace{-0.1cm}\label{eq:geom_dis_PDF_UAS1_LoS_thm}
\end{equation}
}{\small \par}

\vspace{-0.2cm}

\noindent and

\vspace{-0.2cm}

\noindent {\small{}
\begin{equation}
f_{R,n}^{{\rm {NL}}}\left(r\right)=\exp\left(\hspace{-0.1cm}-\hspace{-0.1cm}\int_{0}^{r_{2}}{\rm {Pr}}^{{\rm {L}}}\left(\sqrt{u^{2}+L^{2}}\right)2\pi u\lambda du\right)\exp\left(\hspace{-0.1cm}-\hspace{-0.1cm}\int_{0}^{r}\left(1-{\rm {Pr}}^{{\rm {L}}}\left(\sqrt{u^{2}+L^{2}}\right)\right)2\pi u\lambda du\right)\left(1-{\rm {Pr}}_{n}^{{\rm {L}}}\left(\sqrt{r^{2}+L^{2}}\right)\right)2\pi r\lambda,\label{eq:geom_dis_PDF_UAS1_NLoS_thm}
\end{equation}
}{\small \par}

\noindent {\small{}where $r_{1}$ and $r_{2}$ are given implicitly
by the following equations as}%
{\small \par}

\noindent {\small{}
\begin{equation}
r_{1}=\underset{r_{1}}{\arg}\left\{ \zeta^{{\rm {NL}}}\left(\sqrt{r_{1}^{2}+L^{2}}\right)=\zeta_{n}^{{\rm {L}}}\left(\sqrt{r^{2}+L^{2}}\right)\right\} ,\label{eq:def_r_1}
\end{equation}
}{\small \par}

\noindent \vspace{-0.2cm}
and

\vspace{-0.2cm}

\noindent {\small{}
\begin{equation}
r_{2}=\underset{r_{2}}{\arg}\left\{ \zeta^{{\rm {L}}}\left(\sqrt{r_{2}^{2}+L^{2}}\right)=\zeta_{n}^{{\rm {NL}}}\left(\sqrt{r^{2}+L^{2}}\right)\right\} .\label{eq:def_r_2}
\end{equation}
}{\small \par}

\noindent {\small{}In addition, ${\rm {Pr}}\left[\frac{P\zeta_{n}^{{\rm {L}}}\left(\sqrt{r^{2}+L^{2}}\right)h}{I_{{\rm {agg}}}+N_{0}}>\gamma\right]$
and ${\rm {Pr}}\left[\frac{P\zeta_{n}^{{\rm {NL}}}\left(\sqrt{r^{2}+L^{2}}\right)h}{I_{{\rm {agg}}}+N_{0}}>\gamma\right]$
are respectively computed by }{\small \par}

\noindent {\small{}
\begin{equation}
{\rm {Pr}}\left[\frac{P\zeta_{n}^{{\rm {L}}}\left(\sqrt{r^{2}+L^{2}}\right)h}{I_{{\rm {agg}}}+N_{0}}>\gamma\right]=\exp\left(-\frac{\gamma N_{0}}{P\zeta_{n}^{{\rm {L}}}\left(\sqrt{r^{2}+L^{2}}\right)}\right)\mathscr{L}_{I_{{\rm {agg}}}}^{{\rm {L}}}\left(\frac{\gamma}{P\zeta_{n}^{{\rm {L}}}\left(\sqrt{r^{2}+L^{2}}\right)}\right),\label{eq:Pr_SINR_req_UAS1_LoS_thm}
\end{equation}
}{\small \par}

\noindent {\small{}where $\mathscr{L}_{I_{{\rm {agg}}}}^{{\rm {L}}}\left(s\right)$
is the Laplace transform of $I_{{\rm {agg}}}$ for LoS signal transmission
evaluated at $s$, which can be further written as}{\small \par}

\noindent {\small{}
\begin{equation}
\mathscr{L}_{I_{{\rm {agg}}}}^{{\rm {L}}}\left(s\right)=\exp\left(-2\pi\lambda\int_{r}^{+\infty}\frac{{\rm {Pr}}^{{\rm {L}}}\left(\sqrt{u^{2}+L^{2}}\right)u}{1+\left(sP\zeta^{{\rm {L}}}\left(\sqrt{u^{2}+L^{2}}\right)\right)^{-1}}du\right)\exp\left(-2\pi\lambda\int_{r_{1}}^{+\infty}\frac{\left[1-{\rm {Pr}}^{{\rm {L}}}\left(\sqrt{u^{2}+L^{2}}\right)\right]u}{1+\left(sP\zeta^{{\rm {NL}}}\left(\sqrt{u^{2}+L^{2}}\right)\right)^{-1}}du\right),\label{eq:laplace_term_LoS_UAS1_general_seg_thm}
\end{equation}
}{\small \par}

\noindent \vspace{-0.1cm}
{\small{}and}{\small \par}

\vspace{-0.1cm}

\noindent {\small{}
\begin{equation}
{\rm {Pr}}\left[\frac{P\zeta_{n}^{{\rm {NL}}}\left(\sqrt{r^{2}+L^{2}}\right)h}{I_{{\rm {agg}}}+N_{0}}>\gamma\right]=\exp\left(-\frac{\gamma N_{0}}{P\zeta_{n}^{{\rm {NL}}}\left(\sqrt{r^{2}+L^{2}}\right)}\right)\mathscr{L}_{I_{{\rm {agg}}}}^{{\rm {NL}}}\left(\frac{\gamma}{P\zeta_{n}^{{\rm {NL}}}\left(\sqrt{r^{2}+L^{2}}\right)}\right),\label{eq:Pr_SINR_req_UAS1_NLoS_thm}
\end{equation}
}{\small \par}

\noindent {\small{}where $\mathscr{L}_{I_{{\rm {agg}}}}^{{\rm {NL}}}\left(s\right)$
is the Laplace transform of $I_{{\rm {agg}}}$ for NLoS signal transmission
evaluated at $s$, which can be further written as}{\small \par}

\noindent {\small{}
\begin{equation}
\mathscr{L}_{I_{{\rm {agg}}}}^{{\rm {NL}}}\left(s\right)=\exp\left(-2\pi\lambda\int_{r_{2}}^{+\infty}\frac{{\rm {Pr}}^{{\rm {L}}}\left(\sqrt{u^{2}+L^{2}}\right)u}{1+\left(sP\zeta^{{\rm {L}}}\left(\sqrt{u^{2}+L^{2}}\right)\right)^{-1}}du\right)\exp\left(-2\pi\lambda\int_{r}^{+\infty}\frac{\left[1-{\rm {Pr}}^{{\rm {L}}}\left(\sqrt{u^{2}+L^{2}}\right)\right]u}{1+\left(sP\zeta^{{\rm {NL}}}\left(\sqrt{u^{2}+L^{2}}\right)\right)^{-1}}du\right).\label{eq:laplace_term_NLoS_UAS1_general_seg_thm}
\end{equation}
}{\small \par}
\end{thm}
\begin{IEEEproof}
We omit the proof here due to the page limitation. We will provide
the full proof in the journal version of this paper.
\end{IEEEproof}
\end{algorithm*}

\vspace{-0.4cm}

According to~\cite{our_GC_paper_2015_HPPP,our_work_TWC2016}, we
also investigate the ASE in $\textrm{bps/Hz/km}^{2}$ for a given
$\lambda$, which can be computed as

\begin{singlespace}
\noindent
\begin{equation}
A^{\textrm{ASE}}\left(\lambda,\gamma_{0}\right)=\lambda\int_{\gamma_{0}}^{+\infty}\log_{2}\left(1+\gamma\right)f_{\mathit{\Gamma}}\left(\lambda,\gamma\right)d\gamma,\label{eq:ASE_def}
\end{equation}

\end{singlespace}

\noindent where $\gamma_{0}$ is the minimum working SINR for the
considered SCN, and $f_{\mathit{\Gamma}}\left(\lambda,\gamma\right)$
is the probability density function (PDF) of the SINR observed at
the typical UE at a particular value of $\lambda$. %
Based on the definition of $p^{\textrm{cov}}\left(\lambda,\gamma\right)$
in (\ref{eq:Coverage_Prob_def}), which is the complementary cumulative
distribution function (CCDF) of SINR, $f_{\mathit{\Gamma}}\left(\lambda,\gamma\right)$
can be expressed by

\begin{singlespace}
\noindent
\begin{equation}
f_{\mathit{\Gamma}}\left(\lambda,\gamma\right)=\frac{\partial\left(1-p^{\textrm{cov}}\left(\lambda,\gamma\right)\right)}{\partial\gamma},\label{eq:cond_SINR_PDF}
\end{equation}

\end{singlespace}

\noindent where $p^{\textrm{cov}}\left(\lambda,\gamma\right)$ is
obtained from Theorem~\ref{thm:p_cov_UAS1}.

Considering the results of $p^{\textrm{cov}}\left(\lambda,\gamma\right)$
and $A^{\textrm{ASE}}\left(\lambda,\gamma_{0}\right)$ respectively
shown in (\ref{eq:Theorem_1_p_cov}) and (\ref{eq:ASE_def}), we propose
Theorem~\ref{thm:the_ASE_Falls_Theorem} to theoretically explain
the fundamental reasons of %
\emph{the ASE Crash} discussed in Section~\ref{sec:Introduction}.
\begin{thm}
{\small{}\label{thm:the_ASE_Falls_Theorem}}%
If $L>0$ and $\gamma,\gamma_{0}<+\infty$, then $\underset{\lambda\rightarrow+\infty}{\lim}p^{{\rm {cov}}}\left(\lambda,\gamma\right)=0$
and $\underset{\lambda\rightarrow+\infty}{\lim}A^{{\rm {ASE}}}\left(\lambda,\gamma_{0}\right)=0$.
\end{thm}
\vspace{-0.1cm}

\begin{IEEEproof}
We omit the proof here due to the page limitation. Instead, in the
following, we describe the essence of theorem and provide a toy example
to clarify it. We will provide the full proof in the journal version
of this paper.
\end{IEEEproof}

In essence, Theorem~\ref{thm:the_ASE_Falls_Theorem} states that
when $\lambda$ is extremely large, e.g., in UD SCNs, both $p^{\textrm{cov}}\left(\lambda,\gamma\right)$
and $A^{\textrm{ASE}}\left(\lambda,\gamma_{0}\right)$ will decrease
towards zero with the network densification, and UEs will experience
service outage, thus creating \emph{the ASE Crash}. The fundamental
reason for this phenomenon is revealed by the key point of the proof,
i.e., the signal power will lose its superiority over the interference
power when $\lambda\rightarrow+\infty$, even if the interference
created by the BSs that are relatively far away is ignored.\textbf{
This is because the absolute antenna height difference $L$ introduces}\textbf{\emph{
a cap}}\textbf{ on the signal-link distance and thus on the signal
power.} Theorem~\ref{thm:the_ASE_Falls_Theorem} is in stark contrast
with the conclusion in~\cite{Jeff2011,related_work_Jeff,Related_work_Health,our_GC_paper_2015_HPPP,our_work_TWC2016},
which indicates that the increase in the interference power will be
exactly counter-balanced by the increase in the signal power when
$\lambda\rightarrow+\infty$.

Since the proof of Theorem~\ref{thm:the_ASE_Falls_Theorem} is mathematically
intense and difficult to digest, in the following we provide a toy
example to shed some valuable insights on the rationale behind Theorem~\ref{thm:the_ASE_Falls_Theorem}.
We consider a simple 2-BS SCN as illustrated in Fig.~\ref{fig:model_toy_example_2BSs},
where the 2D distance between the serving BS and the UE and that between
an arbitrary interfering BS and the UE are denoted by $r$ and $\tau r,\left(1<\tau<+\infty\right)$,
respectively. In this example, when $\lambda\rightarrow+\infty$,
then $r\rightarrow0$, which can be intuitively explained by the fact
that the per-BS coverage area is roughly in the order of $\frac{1}{\lambda}$,
and thus the typical 2D distance from the serving BS to the UE approaches
zero when $\lambda\rightarrow+\infty$.

\vspace{-0.4cm}

\noindent \begin{center}
\begin{figure}[H]
\noindent \begin{centering}
\includegraphics[width=8cm]{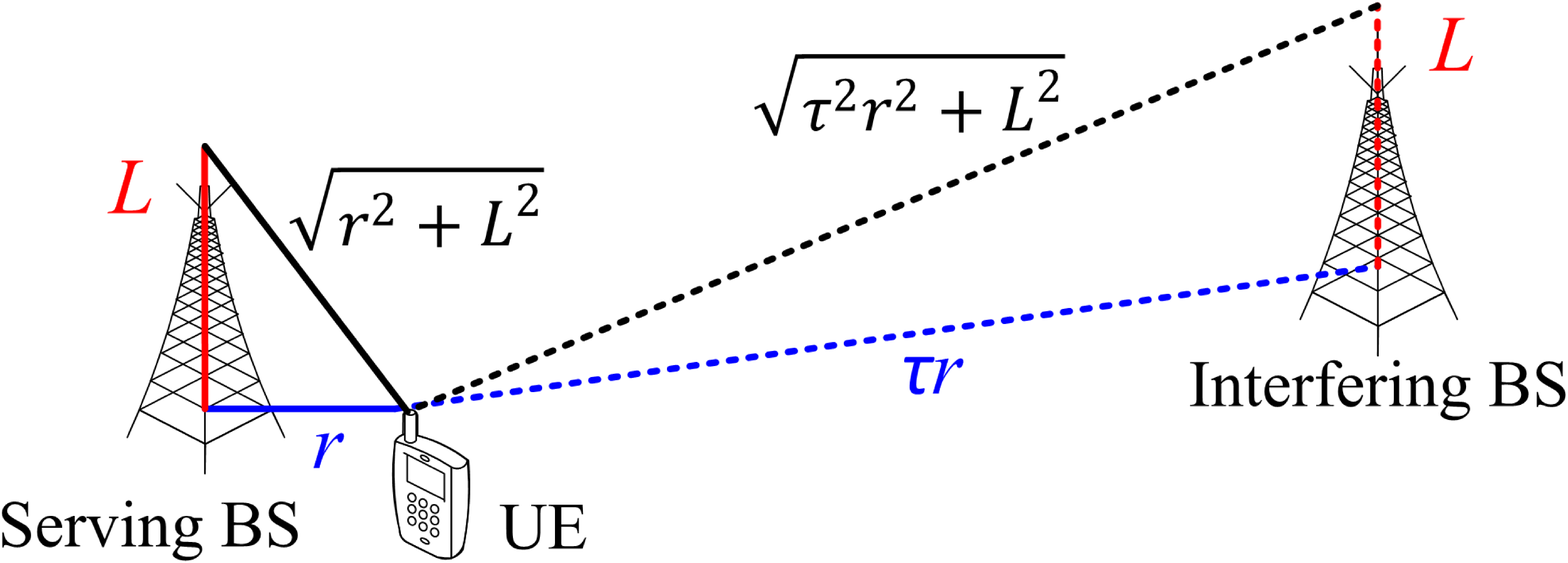}\renewcommand{\figurename}{Fig.}\caption{\label{fig:model_toy_example_2BSs}Illustration of a toy example with
a 2-BS SCN.}
\par\end{centering}
\vspace{-0.5cm}
\end{figure}
\par\end{center}

\vspace{-0.4cm}

Considering that $r\rightarrow0$ and $L$ is smaller than $d_{1}$
in practical SCNs~\cite{SCM_pathloss_model,TR36.828}, we can assume
that both the signal link and the interference link should be dominantly
characterized by \emph{the first-piece LoS path loss function} in
(\ref{eq:PL_BS2UE}), i.e., $\zeta_{1}^{\textrm{L}}\left(w\right)=A_{1}^{{\rm {L}}}\left(\sqrt{r^{2}+L^{2}}\right)^{-\alpha_{1}^{{\rm {L}}}}$.
Thus, based on the 3D distances, we can obtain the signal-to-interference
ratio (SIR) as

\begin{singlespace}
\noindent
\begin{equation}
\bar{\gamma}=\frac{A_{1}^{{\rm {L}}}\left(\sqrt{r^{2}+L^{2}}\right)^{-\alpha_{1}^{{\rm {L}}}}}{A_{1}^{{\rm {L}}}\left(\sqrt{\tau^{2}r^{2}+L^{2}}\right)^{-\alpha_{1}^{{\rm {L}}}}}=\left(\sqrt{\frac{1}{1+\frac{\tau^{2}-1}{1+\frac{L^{2}}{r^{2}}}}}\right)^{-\alpha_{1}^{{\rm {L}}}}.\label{eq:SIR_toy_example}
\end{equation}

\end{singlespace}

\noindent Note that $\bar{\gamma}$ is a monotonically decreasing
function as $r$ decreases when $L>0$. Moreover, it is easy to show
that

\noindent
\begin{equation}
\underset{\lambda\rightarrow+\infty}{\lim}\bar{\gamma}=\underset{r\rightarrow0}{\lim}\bar{\gamma}=\begin{cases}
\begin{array}{l}
1,\\
\tau^{\alpha_{1}^{{\rm {L}}}},
\end{array} & \hspace{-0.2cm}\begin{array}{l}
\left(L>0\right)\\
\left(L=0\right)
\end{array}.\end{cases}\label{eq:limit_SIR}
\end{equation}

Assuming that $\tau=10$ and $\alpha_{1}^{{\rm {L}}}=2$ in (\ref{eq:limit_SIR}),
the limit of $\bar{\gamma}$ in UD SCNs will plunge from 20\,dB when
$L=0$ to 0\,dB when $L>0$, which means that even a rather weak
interferer, e.g., with a power 20\,dB below the signal power, will
become a real threat to the signal link when the absolute antenna
height difference $L$ is non-zero in UD SCNs. The drastic crash of
$\bar{\gamma}$ when $L>0$ is due to \emph{the cap imposed on the
signal power} as the signal-link distance $\sqrt{r^{2}+L^{2}}$ in
the numerator of (\ref{eq:SIR_toy_example}) cannot go below $L$.
Such cap on the signal-link distance and the signal power leads to
\emph{the ASE Crash,} since %
other signal-power-comparable interferers also approach the UE from
all directions as $\lambda$ increases, which will eventually cause
service outage to the UE. %

To sum up, in an UD SCN with conventional deployment (i.e., $L>0$),
both $p^{\textrm{cov}}\left(\lambda,\gamma\right)$ and $A^{\textrm{ASE}}\left(\lambda,\gamma_{0}\right)$
will plunge toward zero as $\lambda$ increases, causing \emph{the
ASE Crash}. Its fundamental reason is \emph{the cap on the signal
power} because of the minimum signal-link distance tied to $L$, which
cannot be overcome with the densification. The only way to avoid \emph{the
ASE Crash} is to remove the signal power cap by setting $L$ to zero,
which means lowering the BS antenna height, not just by a few meters,
but straight to the UE antenna height. Other applicable solutions
may be the usage of very directive antennas and/or the usage of sophisticated
idle modes at the SCN BSs, which will be investigated in our future
work.

\section{Simulation and Discussion\label{sec:Simulation-and-Discussion}}

In this section, we investigate the network performance and use numerical
results to establish the accuracy of our analysis.

As a special case of Theorem~\ref{thm:p_cov_UAS1}, following~\cite{our_work_TWC2016},
we consider a two-piece %
path loss and a linear LoS probability functions %
defined by the 3GPP~\cite{TR36.828,SCM_pathloss_model}. Specifically,
in the path loss model presented in (\ref{eq:prop_PL_model}), we
use $N=2$, $\zeta_{1}^{\textrm{L}}\left(w\right)=\zeta_{2}^{\textrm{L}}\left(w\right)=A^{{\rm {L}}}w^{-\alpha^{{\rm {L}}}}$,
$\zeta_{1}^{\textrm{NL}}\left(w\right)=\zeta_{2}^{\textrm{NL}}\left(w\right)=A^{{\rm {NL}}}w^{-\alpha^{{\rm {NL}}}}$~\cite{TR36.828}.
And in the LoS probability model shown in (\ref{eq:general_LoS_Pr}),
we use $\textrm{Pr}_{1}^{\textrm{L}}\left(w\right)=1-\frac{w}{d_{1}}$
and $\textrm{Pr}_{2}^{\textrm{L}}\left(w\right)=0$, where $d_{1}$
is a constant~\cite{SCM_pathloss_model}. For clarity, this 3GPP
special case %
is referred to as 3GPP Case~1%
. As justified in~\cite{our_work_TWC2016}, we use 3GPP Case~1 for
the case study because it provides tractable results for %
(\ref{eq:geom_dis_PDF_UAS1_LoS_thm})-(\ref{eq:laplace_term_NLoS_UAS1_general_seg_thm})
in Theorem~\ref{thm:p_cov_UAS1}%
. The details are %
relegated to the journal version of this paper.%

Following~\cite{our_work_TWC2016}%
, we adopt the following parameters for 3GPP Case~1%
: $d_{1}=300$\ m, $\alpha^{{\rm {L}}}=2.09$, $\alpha^{{\rm {NL}}}=3.75$,
$A^{{\rm {L}}}=10^{-10.38}$, $A^{{\rm {NL}}}=10^{-14.54}$, $P=24$\ dBm,
$N_{0}=-95$\ dBm%
. %
The BS antenna and the UE antenna heights are set to 10\,m and 1.5\,m,
respectively~\cite{TR36.814}, thus $L=\left|10-1.5\right|=8.5$\,m.%

To check the impact of different path loss models on our conclusions,
we have also investigated the results for a single-slope path loss
model that does not differentiate LoS and NLoS transmissions~\cite{Jeff2011},
where only one path loss exponent $\alpha$ is defined, the value
of which is assumed to be $\alpha=\alpha^{{\rm {NL}}}=3.75$.

\subsection{Validation of Theorem~\ref{thm:p_cov_UAS1} on the Coverage Probability\label{subsec:Sim-p-cov-3GPP-Case-1}}

\noindent \begin{center}
\vspace{-0.4cm}
\begin{figure}
\noindent \begin{centering}
\includegraphics[width=8cm]{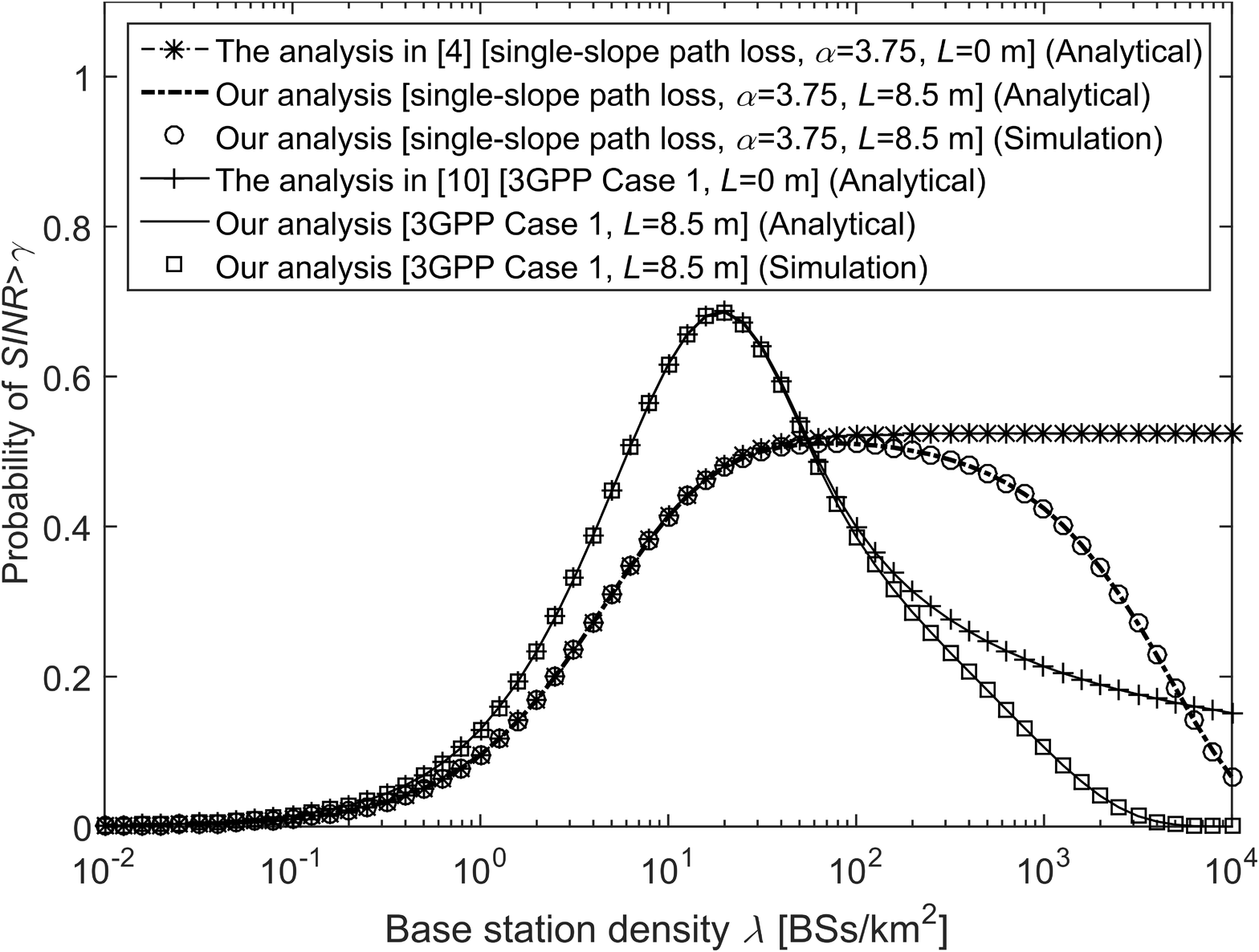}\renewcommand{\figurename}{Fig.}\caption{\label{fig:p_cov_linear_3Dmod_fixedPower24dBm_gamma0dB_wBaseline}$p^{\textrm{cov}}\left(\lambda,\gamma\right)$
vs. $\lambda$ with $\gamma=0\,\textrm{dB}$.}
\par\end{centering}
\vspace{-0.5cm}
\end{figure}
\par\end{center}

\vspace{-0.4cm}

In Fig.~\ref{fig:p_cov_linear_3Dmod_fixedPower24dBm_gamma0dB_wBaseline},
we show the results of $p^{\textrm{cov}}\left(\lambda,\gamma\right)$
with $\gamma=0\,\textrm{dB}$. As can be observed from Fig.~\ref{fig:p_cov_linear_3Dmod_fixedPower24dBm_gamma0dB_wBaseline},
our analytical results given by Theorem~\ref{thm:p_cov_UAS1} match
the simulation results very well, which validates the accuracy of
our theoretical analysis. %
From Fig.~\ref{fig:p_cov_linear_3Dmod_fixedPower24dBm_gamma0dB_wBaseline},
we can draw the following observations which are inline with our discussion
in Section~\ref{sec:Introduction}:
\begin{itemize}
\item For the single-slope path loss model with $L=0$\,m, \emph{the BS
density does NOT matter}, since the coverage probability approaches
a constant for UD SCNs~\cite{Jeff2011}%
.
\item For the 3GPP Case~1 path loss model with $L=0$\,m, \emph{the BS
density DOES matter,} since that coverage probability will decrease
as $\lambda$ increases when the network is dense enough, e.g., $\lambda>20\,\textrm{BSs/km}^{2}$,
due to the transition of a large number of interference paths from
NLoS to LoS~\cite{our_work_TWC2016}. When $\lambda$ is tremendously
large, e.g., $\lambda\geq10^{3}\,\textrm{BSs/km}^{2}$, the coverage
probability decreases at a slower pace because both the interference
and the signal powers are LoS dominated, and thus the coverage probability
approaches a constant related to $\alpha^{{\rm {L}}}$~\cite{Jeff2011,our_work_TWC2016}.
\item For both path loss models, when $L=8.5$\,m, the coverage probability
shows a determined trajectory toward zero in UD SCNs due to \emph{the
cap on the signal power} introduced by the non-zero $L$ as explained
in Theorem~\ref{thm:the_ASE_Falls_Theorem}. In more detail, for
the 3GPP Case~1 path loss model with $\lambda=10^{4}\thinspace\textrm{BSs/km}^{2}$,
the coverage probability decreases from 0.15 when $L=0$\,m to around
$10^{-5}$ when $L=8.5$\,m.
\end{itemize}

\subsection{The Theoretical Results of the ASE\label{subsec:Sim-ASE-3GPP-Case-1}}

\noindent \begin{center}
\vspace{-0.4cm}
\begin{figure}
\noindent \begin{centering}
\includegraphics[width=8cm]{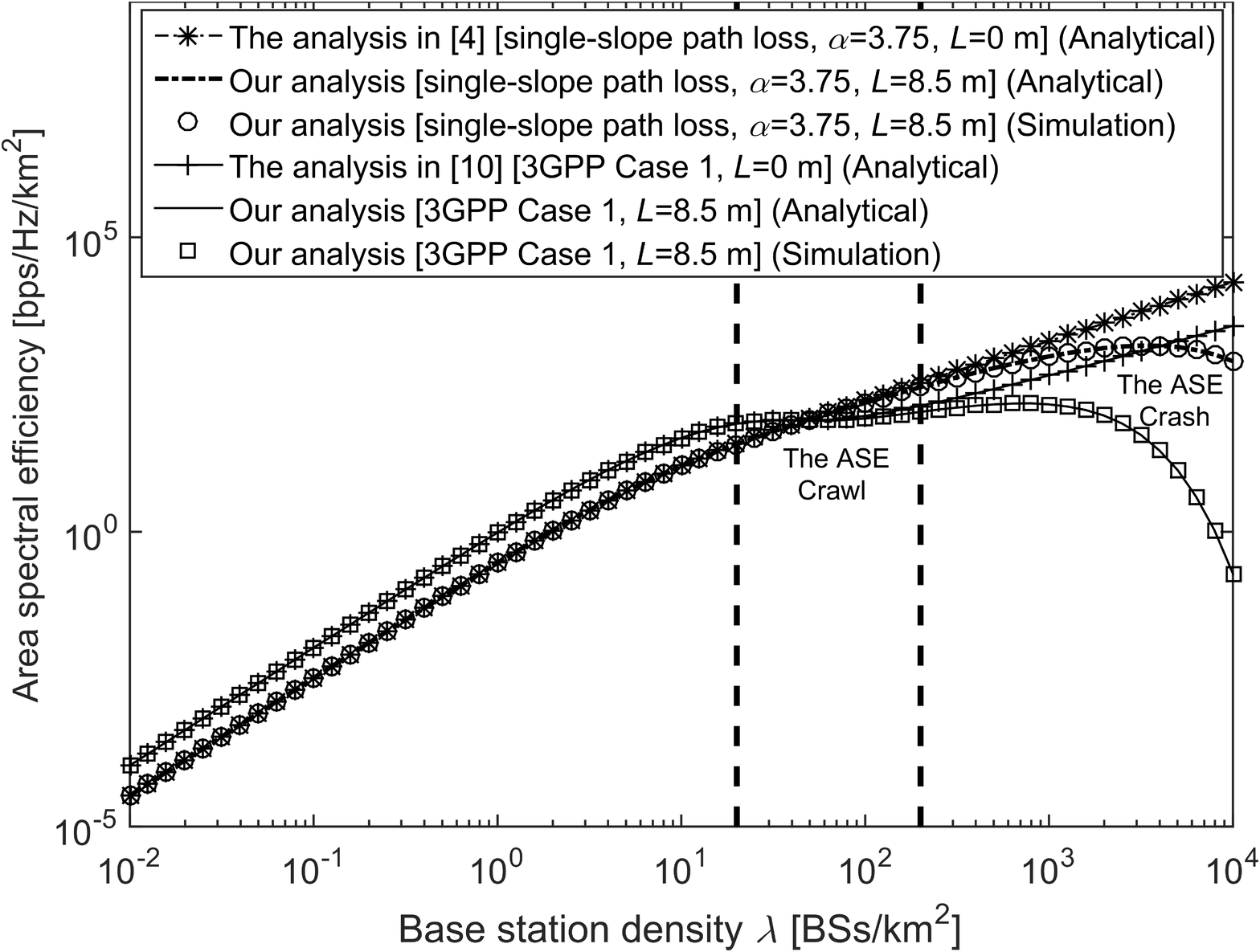}\renewcommand{\figurename}{Fig.}\caption{\label{fig:ASE_linear_3Dmod_fixedPower24dBm_gamma0dB_wBaseline}$A^{\textrm{ASE}}\left(\lambda,\gamma_{0}\right)$
vs. $\lambda$ with $\gamma_{0}=0\,\textrm{dB}$.}
\par\end{centering}
\vspace{-0.5cm}
\end{figure}
\par\end{center}

\vspace{-0.4cm}

In Fig.~\ref{fig:ASE_linear_3Dmod_fixedPower24dBm_gamma0dB_wBaseline},
we show the results of $A^{\textrm{ASE}}\left(\lambda,\gamma_{0}\right)$
with $\gamma_{0}=0\,\textrm{dB}$. %
Fig.~\ref{fig:ASE_linear_3Dmod_fixedPower24dBm_gamma0dB_wBaseline}
is essentially the same as Fig.~\ref{fig:comp_ASE_2Gto5G} with the
same marker styles, except that the results for the single-slope path
loss model with $L=8.5$\,m are also plotted. From Fig.~\ref{fig:ASE_linear_3Dmod_fixedPower24dBm_gamma0dB_wBaseline},
we can confirm the key observations presented in Section~\ref{sec:Introduction}:
\begin{itemize}
\item For the single-slope path loss model with $L=0$\,m, the ASE performance
scales linearly with $\lambda$~\cite{Jeff2011}. The result is promising,
but it might not be the case in reality.
\item For the 3GPP Case~1 path loss model with $L=0$\,m, the ASE suffers
from a slow growth or even a small\emph{ decrease} when $\lambda\in\left[20,200\right]\,\textrm{BSs/km}^{2}$,
i.e., \emph{the ASE Crawl}~\cite{our_work_TWC2016}. %
After \emph{the ASE Crawl}, the ASE grows almost linearly again %
as the network further evolves to an UD one, e.g., $\lambda>10^{3}\thinspace\textrm{BSs/km}^{2}$~\cite{our_work_TWC2016}.
\item For both path loss models with $L=8.5$\,m, the ASE suffers from
severe performance loss in UD SCNs due to \emph{the ASE Crash}, as
explained in Theorem~\ref{thm:the_ASE_Falls_Theorem}. In more detail,
for the 3GPP Case~1 path loss model with $\lambda=10^{4}\thinspace\textrm{BSs/km}^{2}$,
the ASE dramatically decreases from $3141\thinspace\textrm{bps/Hz/km}^{2}$
when $L=0$\,m to $0.2\thinspace\textrm{bps/Hz/km}^{2}$ when $L=8.5$\,m.
\end{itemize}
\textbf{}%

\subsection{Factors that May Impact the ASE Crash\label{subsec:factors_impact_ASE_falls}}

There are several factors that may have large impacts on the existence/severity
of \emph{the ASE Crash}, e.g., various values of $L$ and $\alpha^{{\rm {L}}}$,
Rician fading, etc. In Fig.~\ref{fig:misc_ASE_3Dmod_fixedPower24dBm_gamma0dB},
we investigate the performance of $A^{\textrm{ASE}}\left(\lambda,\gamma_{0}\right)$
for 3GPP Case~1 under the assumptions of $L=3.5$\,m~\cite{TR36.872}
or $\alpha^{{\rm {L}}}=1.09$~\cite{our_work_TWC2016} or Rician
fading~\cite{SCM_pathloss_model}\footnote{Note that here we adopt a practical model of Rician fading in~\cite{SCM_pathloss_model},
where the $K$ factor in dB scale (the ratio between the power in
the direct path and the power in the other scattered paths) is modeled
as $K[dB]=13-0.03w$, where $w$ is the 3D distance in meter. }. Due to the significant accuracy of our analysis, we only show analytical
results of $A^{\textrm{ASE}}\left(\lambda,\gamma_{0}\right)$ in Fig.~\ref{fig:misc_ASE_3Dmod_fixedPower24dBm_gamma0dB}.

Our key conclusions are summarized as follows:
\begin{itemize}
\item Decreasing $L$ from 8.5\,m to 3.5\,m (BS antenna height being 5\,m)
helps to alleviate, but cannot remove \emph{the ASE Crash} unless
$L=0$, as explained in Theorem~\ref{thm:the_ASE_Falls_Theorem}.
From Fig.~\ref{fig:misc_ASE_3Dmod_fixedPower24dBm_gamma0dB}, the
ASE with $L=3.5$\,m peaks at around $\lambda=3000\thinspace\textrm{BSs/km}^{2}$,
but it still suffers from a 60\,\% loss compared with the ASE with
$L=0$\,m at that BS density.
\item Decreasing $\alpha^{{\rm {L}}}$ helps to alleviate \emph{the ASE
Crash} because it softens the SIR crash in (\ref{eq:limit_SIR}).
However, it aggravates \emph{the ASE Crawl} by showing an obvious
ASE decrease when $\lambda\hspace{-0.1cm}\in\hspace{-0.1cm}\left[20,80\right]\,\textrm{BSs/km}^{2}$
due to the drastic interference transition from NLoS to stronger LoS
with $\alpha^{{\rm {L}}}\hspace{-0.1cm}=\hspace{-0.1cm}1.09$~\cite{our_work_TWC2016}.
\item \textcolor{black}{From the simulation results, we can see that Rician
fading makes }\emph{the ASE Crash}\textcolor{black}{{} worse, which
}takes effect early from around $\lambda=400\thinspace\textrm{BSs/km}^{2}$.
The intuition is that \textcolor{black}{the randomness in channel
fluctuation associated with Rician fading is much weaker than that
associated with Rayleigh fading due to the large }$K$ factor in UD
SCNs~\cite{SCM_pathloss_model}. With \textcolor{black}{Rayleigh
fading, some UE in outage might be opportunistically saved by favorable
channel fluctuation of the signal power, while with Rician fading,
such outage case becomes more deterministic due to lack of channel
variation, thus leading to a severer }\emph{ASE Crash}.\textcolor{black}{{}
}%
\end{itemize}
\noindent \begin{center}
\vspace{-0.4cm}
\begin{figure}
\noindent \begin{centering}
\includegraphics[width=8cm]{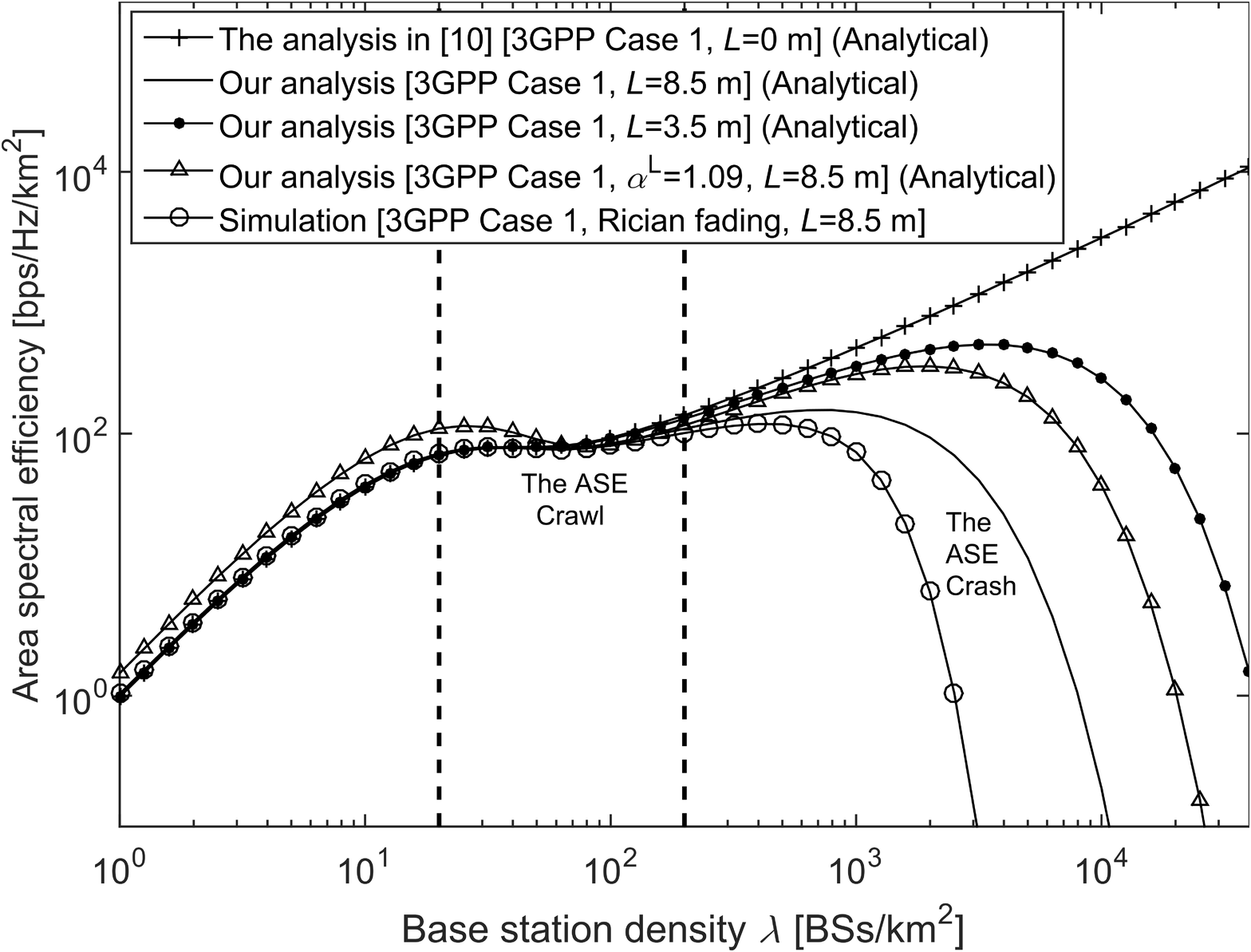}\renewcommand{\figurename}{Fig.}\caption{\label{fig:misc_ASE_3Dmod_fixedPower24dBm_gamma0dB}$A^{\textrm{ASE}}\left(\lambda,\gamma_{0}\right)$
vs. $\lambda$ with $\gamma_{0}=0\,\textrm{dB}$ and various assumptions.}
\par\end{centering}
\vspace{-0.5cm}
\end{figure}
\par\end{center}

\vspace{-0.4cm}

\balance

\subsection{A Novel BS Deployment with UE-Height Antennas\label{subsec:factors_impact_ASE_falls-1}}

Based on our thought-provoking discovery, we make the following recommendation
to vendors and operators around the world: The SCN BS antenna height
must be lowered to the UE antenna height in 5G UD SCNs, so that the
ASE behavior of such networks would roll back to our previous results
in~\cite{our_GC_paper_2015_HPPP,our_work_TWC2016}, thus avoiding
\emph{the ASE Crash}. Such proposed new BS deployment will allow to
realize the potential gains of UD SCNs, but needs a revolution on
BS architectures and network deployment in 5G. The new R\&D challenges
in this area are:
\begin{enumerate}
\item New BS architectures that are anti-vandalism/anti-theft/anti-hacking
at low-height positions.
\item Measurement campaigns for the UE-height channels. Note that at such
height there is an unusual concentration of objects, such as cars,
foliage, etc.
\item Implications of fast time-variant shadow fading due to random movement
of UE-height objects, e.g., cars. %
\item Terrain-dependent network performance analysis.
\item New inter-BS communications based on ground waves.
\item For macrocell BSs with a large $L$, whose BS antenna height cannot
be lowered to the UE antenna height, the existing network performance
analysis for heterogeneous networks may need to be revisited, since
the interference from the macrocell tier to the SCN tier may have
been greatly over-estimated due to the common assumption of $L=0$.
\end{enumerate}

\section{Conclusion\label{sec:Conclusion}}

We presented a new and significant theoretical discovery, i.e., the
serious problem of \emph{the ASE Crash}. If the absolute height difference
between BS antenna and UE antenna is larger than zero, then the ASE
performance will continuously decrease with network densification
for UD SCNs. The only way to fully overcome \emph{the ASE Crash} is
to lower the SCN BS antenna height to the UE antenna height, which
will revolutionize the approach of BS architecture and deployment
in 5G. In our future work, we will also study how the usage of very
directive antennas and/or the usage of sophisticated idle modes at
the small cell BSs can help to mitigate \emph{the ASE Crash}.

\bibliographystyle{IEEEtran}
\bibliography{Ming_library}

\begin{thebibliography}{10}
\providecommand{\url}[1]{#1}
\csname url@samestyle\endcsname
\providecommand{\newblock}{\relax}
\providecommand{\bibinfo}[2]{#2}
\providecommand{\BIBentrySTDinterwordspacing}{\spaceskip=0pt\relax}
\providecommand{\BIBentryALTinterwordstretchfactor}{4}
\providecommand{\BIBentryALTinterwordspacing}{\spaceskip=\fontdimen2\font plus
\BIBentryALTinterwordstretchfactor\fontdimen3\font minus
  \fontdimen4\font\relax}
\providecommand{\BIBforeignlanguage}[2]{{%
\expandafter\ifx\csname l@#1\endcsname\relax
\typeout{** WARNING: IEEEtran.bst: No hyphenation pattern has been}%
\typeout{** loaded for the language `#1'. Using the pattern for}%
\typeout{** the default language instead.}%
\else
\language=\csname l@#1\endcsname
\fi
#2}}
\providecommand{\BIBdecl}{\relax}
\BIBdecl

\bibitem{Webb_survey}
{ArrayComm \& William Webb}, Ofcom, London, U.K., 2007.

\bibitem{Tutor_smallcell}
D.~López-Pérez, M.~Ding, H.~Claussen, and A.~Jafari, ``{Towards 1 Gbps/UE in
  cellular systems: Understanding ultra-dense small cell deployments},''
  \emph{IEEE Communications Surveys Tutorials}, vol.~17, no.~4, pp. 2078--2101,
  Jun. 2015.

\bibitem{TR36.872}
3GPP, ``{TR 36.872: Small cell enhancements for E-UTRA and E-UTRAN - Physical
  layer aspects},'' Dec. 2013.

\bibitem{Jeff2011}
J.~Andrews, F.~Baccelli, and R.~Ganti, ``A tractable approach to coverage and
  rate in cellular networks,'' \emph{IEEE Transactions on Communications},
  vol.~59, no.~11, pp. 3122--3134, Nov. 2011.

\bibitem{TR36.828}
3GPP, ``{TR 36.828: Further enhancements to LTE Time Division Duplex (TDD) for
  Downlink-Uplink (DL-UL) interference management and traffic adaptation},''
  Jun. 2012.

\bibitem{SCM_pathloss_model}
{Spatial Channel Model AHG}, ``{Subsection 3.5.3, Spatial Channel Model Text
  Description V6.0},'' Apr. 2003.

\bibitem{related_work_Jeff}
X.~Zhang and J.~Andrews, ``Downlink cellular network analysis with multi-slope
  path loss models,'' \emph{IEEE Transactions on Communications}, vol.~63,
  no.~5, pp. 1881--1894, May 2015.

\bibitem{Related_work_Health}
T.~Bai and R.~Heath, ``Coverage and rate analysis for millimeter-wave cellular
  networks,'' \emph{IEEE Transactions on Wireless Communications}, vol.~14,
  no.~2, pp. 1100--1114, Feb. 2015.

\bibitem{our_GC_paper_2015_HPPP}
M.~Ding, D.~López-Pérez, G.~Mao, P.~Wang, and Z.~Lin, ``Will the area spectral
  efficiency monotonically grow as small cells go dense?'' \emph{IEEE GLOBECOM
  2015}, pp. 1--7, Dec. 2015.

\bibitem{our_work_TWC2016}
M.~Ding, P.~Wang, D.~López-Pérez, G.~Mao, and Z.~Lin, ``Performance impact of
  {LoS and NLoS} transmissions in dense cellular networks,'' \emph{IEEE
  Transactions on Wireless Communications}, vol.~15, no.~3, pp. 2365--2380,
  Mar. 2016.

\bibitem{TR36.814}
3GPP, ``{TR 36.814: Further advancements for E-UTRA physical layer aspects
  (Release 9)},'' Mar. 2010.

\end{thebibliography}

\end{document}